# Factors Affecting The Intensity of Solar Energetic Particle Events

Nat Gopalswamy

*Solar Physics laboratory, Code 671, NASA Goddard Space Flight Center, Greenbelt, MD 20771, USA*

**Abstract.** This paper updates the influence of environmental and source factors of shocks driven by coronal mass ejections (CMEs) that are likely to influence the solar energetic particle (SEP) events. The intensity variation due to CME interaction reported in [1] is confirmed by expanding the investigation to all the large SEP events of solar cycle 23. The large SEP events are separated into two groups, one associated with CMEs running into other CMEs, and the other with CMEs running into the ambient solar wind. SEP events with CME interaction generally have a higher intensity. New possibilities such as the influence of coronal holes on the SEP intensity are also discussed. For example, the presence of a large coronal hole between a well-connected eruption and the solar disk center may render the shock poorly connected because of the interaction between the CME and the coronal hole. This point is illustrated using the 2004 December 3 SEP event delayed by about 12 hours from the onset of the associated CME. There is no other event at the Sun that can be associated with the SEP onset. This event is consistent with the possibility that the coronal hole interaction influences the connectivity of the CMEs that produce SEPs, and hence the intensity of the SEP event.

## INTRODUCTION

Large solar energetic particle (SEP) events are associated with powerful coronal mass ejections (CMEs). The SEPs are thought to be accelerated by shocks driven by these CMEs (see e.g., [2]). Faster CMEs drive stronger shocks, so one expects a good correlation between CME speed and particle intensity. The correlation between SEP intensity and CME speed has been less than perfect: for a given CME speed, the SEP intensity can vary over 3 orders of magnitude (see e.g., [3]). The reason for this discrepancy is not fully understood, although several explanations have been proposed: particle acceleration depends on both the source properties such as shock strength and the free energy in the active region, and environmental properties such as the presence of seed particles, inhomogeneous upstream medium due to the presence of preceding CMEs and/or turbulence, orientation of the ambient magnetic field with respect to shock normal, and the connectivity of the acceleration region to the observing spacecraft. Additional possibilities are the presence of coronal holes near the eruption region and the Alfven speed distribution in the ambient medium.

When a CME erupts near a streamer, the shock section passing through the streamer is likely to be stronger compared to the non-streamer regions. This is because of the enhanced density and reduced Alfvén speed in the streamer region. Thus the efficiency of particle acceleration may vary along the shock surface. Similarly, when the shock passes through a tenuous region, the shock may weaken due to higher Alfven speed in the tenuous medium. By studying the ability of fast and wide CMEs in producing type II radio bursts, it has been shown that the Alfven speed in the ambient medium can vary by a factor of 4 from 400 to 1600 km/s [4]. Thus a 400 km/s CME can be super-Alfvénic, while a 1600 km/s CME can be sub-Alfvénic demonstrating that the ambient medium can severely affect the particle acceleration efficiency. It was recently shown that coronal holes can deflect CMEs by large angles (~50 degrees) such that CMEs originating from the disk center behave like limb CMEs [5]. Such a deflection can change the magnetic connectivity of the acceleration region to the observer from a favorable to an unfavorable one or vice versa. This represents another environmental factor, which needs to be studied for a better understanding of the SEP intensity variation. When a coronal hole is present between the Sun – observer line and the solar source of the SEP event, the SEP intensity has been reported to be diminished [6]. However, other investigations [7,8] did not find any bias against SEP production in fast solar wind regions or coronal holes (in which the Alfvén speed is expected to be higher). Thus the influence of coronal holes on SEP events remains a controversial issue. While these authors considered the shock strength in the ambient medium, we focus on the deflection of CMEs by coronal holes. Change in the trajectory of CMEs has been shown to occur due to CME collisions [9]. In this paper, we first confirm the importance of CME interaction with other CMEs and streamers as in [1] using a larger set of data (all SEP events of solar cycle 23 listed by NOAA/SWPC) and present some case studies that clearly suggest the influence of coronal holes.

## PRECEDING CMES AND SEP INTENSITY

Using a set of 60 large SEP events (1996 – 2002) it was found that CMEs that erupt soon after a previous CME from the same active region are much more efficient in accelerating particles than those with no such preceding CMEs within a day [1]. The coronal environment for the second CME is clearly different from the undisturbed corona. Additional particle acceleration due to CME-CME interaction was first discovered in the radio dynamic spectrum of a low-frequency (1 – 2 MHz) type II radio burst that showed a sudden enhancement at the time of interaction between two CMEs [9]. The enhancement requires additional acceleration of electrons, which led to the question of possible increase in particle acceleration efficiency due to the presence of preceding CME.

Figure 1 shows a spectacular CME-CME interaction event in which two fast CMEs (CME1 1000 km/s; CME2 2567 km/s) interacted within the coronagraphic field of view. The radio signature can be seen in the Wind/WAVES radio dynamic spectrum as an enhancement in the interval 21:12 – 21:42 UT as the type II burst from CME2 was in progress. The enhancement seems to happen when the shock producing the type II burst gets very close to CME1. The coronagraphic image at 21:18 UT obtained by the C3 telescope of the Solar and Heliospheric Observatory (SOHO) mission's Large Angle and Spectrometric Coronagraph (LASCO) happens to be right at the beginning of the radio enhancement, thus showing the relative positions of the two

CMEs. The radio enhancement implies additional electrons accelerated at the shock front either due to the same mechanism becoming more efficient or an additional mechanism coming into play. The higher frequency of the radio enhancement with respect to the underlying type II bursts suggests that the local plasma frequency is elevated for the duration of the enhancement. One interpretation is that the shock is

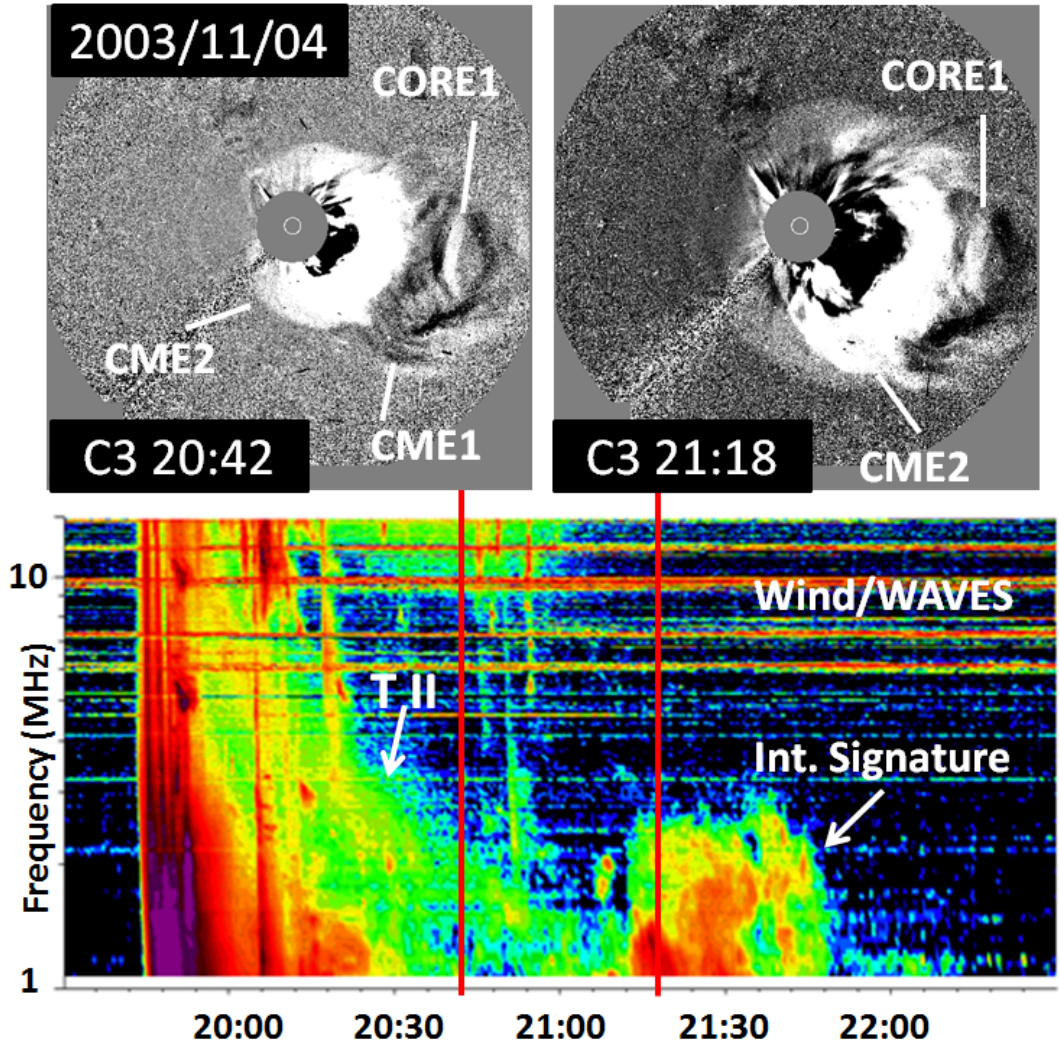

**FIGURE 1.** Interacting CMEs observed by SOHO/LASCO/C3 on 2003 November 4 (top) and the low-frequency interaction radio signature of CME interaction detected by the Wind/WAVES experiment. The times of the coronagraphic images are shown by the vertical lines on the radio dynamic spectrum. The interaction signature and the underlying type II burst (T II) are indicated by arrows.

entering into a higher density medium (provided by the preceding CME) and hence a higher plasma frequency. From the height-time plots of CME1 and CME2, we infer that the radio enhancement begins when CME2 is at a heliocentric distance of 18 solar radii (Rs) and ends when the leading edge is at ~25 Rs. CME2 was also associated with a large SEP event. There may be signatures in the time profile of the SEP event corresponding to the radio enhancement, which need to be further investigated keeping mind the Sun-Earth travel time of the SEPs. We refer to SEP events from

CMEs such as CME2 with a preceding CME within 24 hours as P events. When an SEP event has no such preceding CME, we refer to it an NP event. We now compare the properties of P and NP events that occurred during solar cycle 23.

**TABLE 1.** Comparison between P and NP events of solar cycle 23.

| Property | P Events | NP Events |
|---|---|---|
| Average CME speed (km/s) | 1708 | 1545 |
| Halo CME fraction (%) | 86 | 69 |
| Median SEP intensity | 317 | 35 |
| Active Region area (msh) | 910 | 500 |
| High intensity ($\geq$ 50 pfu) events (%) | 70 | 38 |
| Low intensity (<50 pfu) events (%) | 30 | 62 |

Table 1 summarizes the properties of P and NP type SEP events, and those of the associated CMEs. Kinematic properties of CMEs associated with the P and NP events are very similar: the average CME speeds differ by less than 10% while the halo CME fraction differs by less than 20% (the halo CME fraction is indicative of the average energy of the CME population, see [10]). On the other hand, the median SEP intensity is almost an order of magnitude higher for the P events (317 pfu vs. 35 pfu; pfu = particle flux units, denoting 1 particle per $cm^2$.s.sr). If we define 50 pfu as the dividing line between high and low intensity events, then we see that a majority of the P events fall into the high intensity groups, while the NP events constitute the majority in the low intensity group. The dividing line of 50 pfu is somewhat arbitrary, but represents 5 times the minimum intensity level (10 pfu in the > 10 MeV GOES energy channel) for a large SEP event. Thus Table 1 confirms that the SEP events caused by CMEs with preceding CMEs tend to have higher intensities. The area of the source active region (in units of msh – millionths of the solar hemisphere) for NP events is only half of that for P events. It is known that faster CMEs originate from active regions of larger area [11], so more high speed CMEs may originate from larger active regions thus creating opportunity for CME interaction.

## CORONAL HOLES AND SEP EVENTS

In this section, we present observations related to three SEP events, one of which is a large SEP event, and the other two have intensity in the range 1 – 10 pfu and hence do not qualify as large SEP events (see Table 2). All three events are from the western hemisphere within the longitudinal range from which most of the large SEP events originate [4]. The striking observation is that the SEP onset times show peculiar relationship to the onset times of the CMEs that cause these SEP events. Figure 1 shows the time profiles of the three SEP events and the solar source locations from the Extreme-ultraviolet Imaging Telescope (EIT) on board SOHO and the soft X-ray telescope (SXT) on board Yohkoh. All the three CMEs erupted from source regions near coronal holes. For the 2003 November event (see Fig. 2b), the coronal hole is located to the east of the eruption region, so a deflection would improve the connectivity of this event, for the CME becomes more western. The SEP onset is prompt in this case, consistent with a well-connected event. On the other hand, for the December 2 event the coronal hole was to the west of the eruption region (see Fig.

2d), so the CME would deflect to the east, thereby reducing the connectivity of its shock (the nose moves to the east). Accordingly, the SEP onset is delayed by ~12 hours from the CME onset. For poorly connected events, the delay is generally interpreted as the time required for the shock to encompass the field lines connected to the observer. We note that there is no other CME at the Sun that can be associated with the onset at ~12 UT on 2004 December 3, except for the CME in question. These two events suggest the possibility that the coronal hole interaction influences the connectivity of the CMEs that produce SEPs, and hence the intensity of the SEP event.

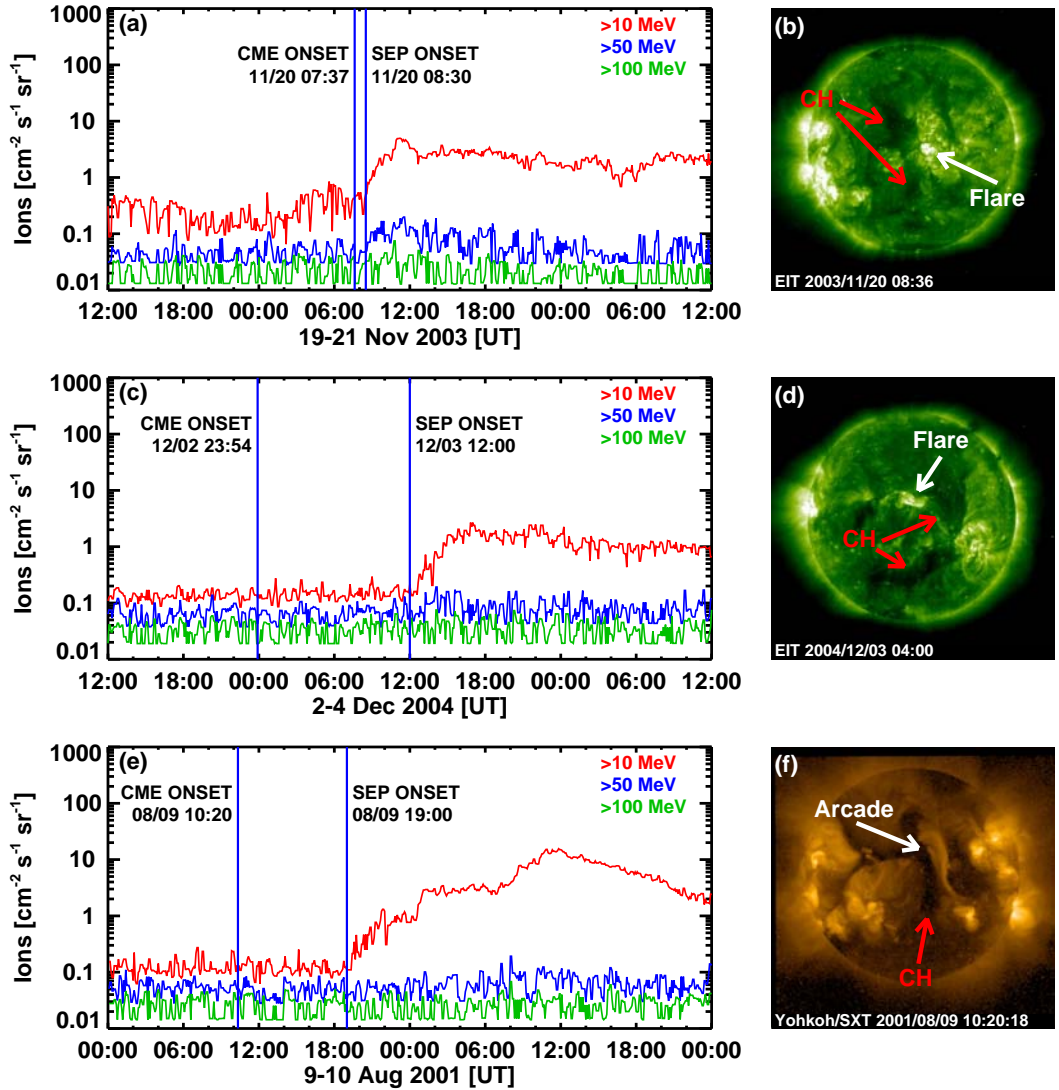

**FIGURE 2.** Three SEP events that have unconventional time profiles and intensity values and the solar sources revealed in EUV or X-ray images. In each case, one or more coronal holes are located close to the eruption region. In the top two cases the eruption is from active regions. In the bottom case is a filament eruption near the coronal hole. Vertical lines mark the SEP and CME onset times.

The 2001 August event (Fig. 2e) was somewhat peculiar in that it was a large SEP event (Intensity >10 pfu) associated with a 479 km/s CME and a C3.7 flare. The event was extremely complex in that the coronal hole disappeared in the aftermath of the eruption and was not seen after the event. The CME also had a preceding CME, so the CME interaction needs to be considered as well. The CME was also rather slow and one of the slowest ones to be associated with a large SEP event. Finally, the filament that erupted in this event was roughly of north-south orientation, so the CME flux rope is expected to be roughly perpendicular to the ecliptic plane. This makes the shock more extended in the latitudinal direction than in the longitudinal direction requiring additional time before the shock encompasses the field lines connected to the observer.

**TABLE 2.** Properties of three SEP events and the associated CMEs shown in Figure 2

| Date | SEP Time | CME Time | Location | X-ray Imp. | CME Speed[b] | SEP intensity[c] |
|---|---|---|---|---|---|---|
| 2003/11/20 | 08:30 | 07:37 | N01W08 | M9.8 | 890 | 5 pfu |
| 2004/12/03 | 12:00 | 23:54[a] | N08W02 | M1.5 | 1216 | 2 pfu |
| 2001/08/09 | 19:00 | 10:20 | N10W15 | C3.7 | 479 | 20 pfu |

[a]Time (UT) corresponds to the previous day. [b]In km/s. [c]Peak value

# SUMMARY

Observations presented in this paper confirm that both source and environmental factors may affect the intensity of SEP events. Statistical studies involving all the SEP events of solar cycle 23 do indicate the importance of CME-CME interaction, which included CME interaction with coronal streamers. Taking clues from a recent discovery that coronal holes change the trajectory of CMEs, we reported a few cases in which the coronal holes seem to affect the magnetic connectivity of SEP source regions to the observer. These preliminary results suggest that we have not exhausted the study of all the environmental factors that might affect the SEP intensity, which is very much needed to fully understand the intensity variation of SEP events.

# ACKNOWLEDGMENTS

The author thanks P. Mäkelä for help with Fig. 2. SOHO is a project of international collaboration between ESA and NASA. Work supported by NASA's LWS/TRT program.